\documentclass[aps,preprint,showpacs,showkeys,floatfix,nofootinbib]{revtex4-1}

\usepackage{dcolumn}

\usepackage{extarrows}
\usepackage{graphicx}
\usepackage{graphicx,epstopdf}
\usepackage{dcolumn}
\usepackage{bm}

\newcommand{\nn}{\nonumber}

\begin{document}

\title[]{Roto-translational Effects on Deflection of Light and Particle by Moving Kerr Black Hole}
\author{Guansheng He and Wenbin Lin}

\email{Email: wl@swjtu.edu.cn. }

\affiliation{
School of Physical Science and Technology, Southwest Jiaotong University, Chengdu 610031, China
}%

\date{\today}

\begin{abstract}
Velocity effects in first-order Schwarzschild deflection of light and particles have been explored in the previous literature. In this paper, we investigate the roto-translational-motion induced deflection by one moving Kerr black hole with an arbitrary but constant speed. It is shown that the coupling between the effects of the rotation and the translational motion always exists for both light and particles. The contribution of the roto-translational deflection to the total bending angle is discussed in detail. This ratio takes upper limit for light and it decreases monotonically with increasing translational velocity for a massive particle. For a given translational velocity of black hole, this ratio increases with the particle' velocity. In addition, the Post-Newtonian dynamics of the photon and particle is also presented.
\end{abstract}


\maketitle

\section{introduction}

Gravitational lensing is one of the most significant focuses for theory of gravitation in astrophysics and cosmology. The deflection of light by a static Schwarzschild or Kerr black hole has been extensively studied, and the calculations have been performed to at least third orders (see, e.g., Refs. \cite{SerenoLuca2006,AazKeePett2011}, and references therein). The motion effects of the black hole on the deflection have also been investigated by several groups. In 2002, Kopeikin and Mashhoon~\cite{KopeiMash2002} studied the second-order purely spin-induced light deflection (gravitomagnetic effects), based on the Lorentz covariant theory for first-order deflection of light by overall lenses in arbitrary motion~\cite{KopeiSch1999}. Sereno~\cite{Sereno2002b,Sereno2005b} calculated the roto-translational effects on deflection by the mass distribution with low velocity via the time delay of light. Wucknitz and Sperhake~\cite{Olaf2004} studied the first-order deflection of both particle and light for a moving Schwarzschild black hole, and their results showed the effects due to the translational motion of lens may bring about strong corrections to the total deflection angle. It can be expected that the kinematic effects will also play an important role in the higher orders.

In this work, we work on the leading roto-translational effects of a moving Kerr black hole on the deflection of both light and particle. As a first step to this complicated issue, we limit our discussions for the case in which the translational velocity and the angular momentum of the Kerr black hole are perpendicular.

This paper is organized as follows. The derivation of the harmonic metric of a uniformly moving Kerr black hole is presented in Section~\ref{harmonic metric}. Section~\ref{bending angle} gives the calculation of deflection of light and particle. Section~\ref{kinematic discussion} is devoted to detailed discussions about the leading roto-translational effects in the weak-field approximation. Section~\ref{discussion} contains the summary and conclusion.

To be specific, we use units where $G=c=1$ throughout.

\section{harmonic metric for moving kerr black hole}\label{harmonic metric}
Let $\bm{e}_i~(i=1,2,3)$ denotes the unit vector of three-dimensional rectangular coordinate system. Suppose Kerr black hole has a mass of $m$ and an angular momentum of $J$ in $\bm{e}_3$ direction. The harmonic metric of Kerr black hole in the center of mass's rest frame can be written as follows~\cite{LJ2013}:
{\small\begin{align}
&\nn ds^{2}=-dX^{2}_{0}+\frac{R^2(R+m)^2\!+\!a^2X^2_3}{\left(R^2\!+\!\frac{a^2}{R^2}X^2_3\right)^2}\!\left[\!\frac{\left(\mathbf{X}\!\cdot\! d\mathbf{X}+\frac{a^2}{R^2}X_3dX_3\right)^{2}}{R^2+a^2-m^2} \!+\!\frac{X_3^2}{R^2} \frac{\left(\mathbf{X}\!\cdot\! d\mathbf{X}-\frac{R^2}{X_3} d X_3\right)^{\!2}}{R^2-X_3^2}\right]\\
&\nn+\frac{(R+m)^{2}+a^{2}}{R^2-X_3^2}\!\!\left[\!\frac{R^2m^2a \! \left(R^2\!-\!X_3^2\right) \! \left(\mathbf{X}\! \cdot \! d\mathbf{X}+\frac{a^2}{R^2}X_3dX_3\right)}{(R^2+a^2-m^2)(R^2+a^2)(R^4+a^2X^2_3)}
\!+\!\frac{R\left(X_2dX_1\!-\!X_1dX_2\right)}{R^2+a^2}\!\right]^2 \\
&+\!\frac{2m(R\!+\!m)}{(R\!+\!m)^2\!+\!\frac{a^2}{R^2}\!X^2_3}\!\!\left[\!\!\frac{Rm^2a^2\!\left(R^2\!\!-\!\!X_3^2\right)\! \left(\!\mathbf{X}\!\cdot\!d\mathbf{X}\!+\!\frac{a^2}{R^2}X_3dX_3\!\right)}{(R^2\!+\!a^2\!-\!m^2)(R^2\!+\!a^2)\left(R^4\!+\!a^2X^2_3\right)}
\!+\!\frac{a\left(X_2dX_1\!-\!X_1dX_2\right)}{R^2+a^2}\!+\!dX_{0}\!\right]^{\!2}\!,  \label{HarmonicKerr}
\end{align}}
where $a \equiv \frac{J}{m}$ is the angular momentum per mass and usually $\left|a\right|\leq m$. $\mathbf{X}\!\cdot\! d\mathbf{X}\equiv X_1dX_1\!+\!X_2dX_2\!+\!X_3dX_3$, and ~$\frac{X_1^2+X_2^2}{R^2+a^2}+\frac{X_3^2}{R^2}=1$.

In the limit of weak field, the harmonic Kerr metric (up to order$~1/R^3$) is reduced to
\begin{eqnarray}
&&g_{00}=-1-2\phi\left(1-\frac{a^2X_3^2}{R^4}\right)-2\phi^2-2\phi^3~, \label{g00s}\\
&&g_{0i}=(1+\phi)\zeta_i~, \label{g0is}\\
&&g_{ij}=\!\left[1-2\phi\left(1-\frac{a^2X_3^2}{R^4}\right)+\phi^2\right]\delta_{ij}-\frac{\phi(\zeta_iX_j+\zeta_jX_i)}{2R}
+(1-2\phi)\phi^2\frac{X_iX_j}{R^2}~, \label{gijs}
\end{eqnarray}
where $\phi\equiv-\frac{m}{R}$ is Newtonian gravitational potential when $R\rightarrow \infty$, and $~\bm{\zeta}\equiv \frac{2am}{R^3}\left(\bm{X} \times \bm{e_3}\right)$~denotes the vector potential due to the rotation of Kerr black hole.

The metric for a uniformly moving Kerr black hole can be obtained from Eqs.~\eqref{g00s}\,-\,\eqref{gijs} via Lorentz transformation.~Suppose the coordinate frame of the background is denoted by $(t,~x,~y,~z)\,$, and the translational velocity of the black hole is assumed to be ${\bm v}=v\bm{e}_1$. The Lorentz transformation between $(t,~x,~y,~z)$ and  $(X_0,~X_1,~X_2,~X_3)$ can be written as
\begin{eqnarray}
&& X_0 = \gamma t-v\gamma x~,\nonumber  \label{LorentzTran1} \\
&& X_1 = \gamma x- v\gamma t~,\nonumber  \label{LorentzTran2} \\
&& X_2 = y~,\nonumber   \label{LorentzTran3} \\
&& X_3 = z~,  \label{LorentzTran4}
\end{eqnarray}
where $\gamma=(1-v^2)^{-\scriptstyle \frac{1}{2}}$ is Lorentz factor. Therefore, the metric of a uniformly moving Kerr black hole in the weak field limit is obtained as follows:
{\small
\begin{eqnarray}
\nn&&g_{00}=\!-1\!-\!2(1\!+\!v^2)\gamma^2\Phi\left(\!1\!-\!\frac{a^2X_3^2}{R^4}\right)\!-\!(1\!+\!\gamma^2)\Phi^2\!-\!2\gamma^2\Phi^3
\!-\!v\gamma^2\!\left[2(1\!+\!\Phi)\!+\!v\frac{X_1}{R}\Phi\right]\zeta_1 \\
&&~~~~~~~+v^2\gamma^2(1-2\Phi)\Phi^2\frac{X_1^2}{R^2}~,\label{g00m}\\
\nn&&g_{0i}=\gamma^{1+\delta_{i1}}(1+\Phi)\zeta_i\!+\!v^2\gamma^2(1+\Phi)\zeta_1\delta_{i1}
\!+\!v\gamma^2\!\left(\!1+2\Phi+2\Phi^2+2\Phi^3-\frac{2\Phi a^2X^2_3}{R^4}\right)\!\delta_{i1}\\
&&~~~~~~~-\!v\gamma^{1+\delta_{i1}}\!\!\left[\!\left(\!1\!-\!2\Phi\!+\!\Phi^2\!+\!\frac{2\Phi a^2X^2_3}{R^4}\!\right)\delta_{i1}
\!-\!\frac{\Phi(\zeta_1X_i\!+\!\zeta_iX_1)}{2R}\!+\!(1\!-\!2\Phi)\Phi^2\frac{X_1X_i}{R^2}\!\right]~,\label{g0im}\\
\nn&&g_{ij}=(\,1-\Phi\,)^2\,\delta_{ij}-v^2\gamma^2\,\Phi\,(\,4+\Phi+2\,\Phi^2\,)\,\delta_{i1}\,\delta_{j1}
+(\,2\,v^2\gamma^2\,\delta_{i1}\,\delta_{j1}+\delta_{ij}\,)\,\frac{2\,\Phi\,a^2X_3^2}{R^4}\\
&&~~~~~~~-\!v\gamma\!\left[\gamma^{\delta_{j1}}\!\delta_{i1}\zeta_j\!+\!\gamma^{\delta_{i1}}\!\delta_{j1}\zeta_i\right]\!(1\!\!+\!\!\Phi)
\!+\!\gamma^{\delta_{i1}\!+\delta_{j1}}\!\!\left[\!(1\!-\!2\Phi)\Phi^2\frac{X_iX_j}{R^2}
\!-\!\frac{\Phi(\zeta_iX_j\!+\!\zeta_jX_i)}{2R}\!\right]~.~~~\label{gijm}
\end{eqnarray}}
It can be verified that Eqs. \eqref{g00m}\,-\,\eqref{gijm} reduce to the post-Newtonian approximation and the multipole fields~\cite{Weinberg1972} in the limit of $v\rightarrow 0$.

\section{deflection of test particle in the observer frame} \label{bending angle}
\begin{figure}
  \centering
  \includegraphics[width=12.5cm]{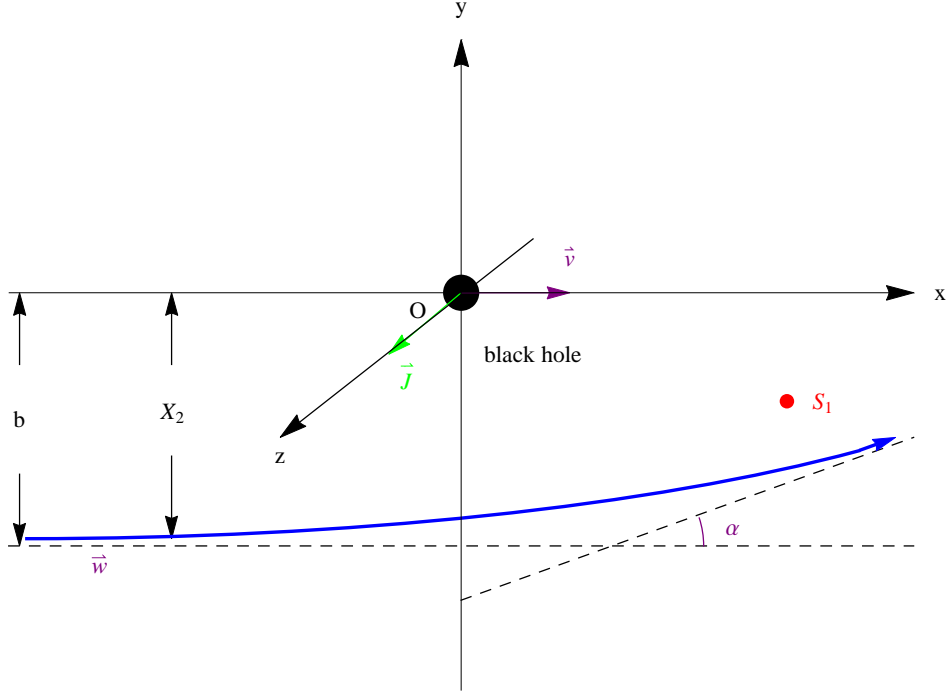}
  \caption{Sketch map of deflection of the test particle by moving Kerr black hole. The deflection angle $~\alpha~$ is greatly exaggerated and defined to be positive. The blue line represents the trajectory of the test particle which comes from $x=-\infty$ with velocity $\bm{w}$. The observer $S_1$ is
  static relative to the background mentioned above. The angular momentum $~\bm{J}~$ of Kerr black hole is along positive z-axis so $~a>0~$ and the test particle takes the prograde motion relative to the rotation of Kerr black hole.}
  \label{figure1}
\end{figure}

The Lagrange function and corresponding Lagrange equation for the test particle (photon or particle) are
\begin{eqnarray}
&&L=\left(\frac{ds}{d\xi}\right)^2~,\\
&& \frac{d}{d\xi}\frac{\partial L}{\partial \overset{.}q}-\frac{\partial L}{\partial q}=0~,
\end{eqnarray}
where $\xi$ is an affine parameter for the test particle's world line, $q$ denotes $t$,~$x$,~$y$,~or $z$, and the line element $ds^2$ is dictated by Eqs. \eqref{g00m}\,-\,\eqref{gijm}. For simplicity, we only consider the motion on the equatorial plane of black hole, namely $x-y$~plane. The velocity of test particle (or light) is denoted as $\bm{w}$ with $\bm{w}|_{x=-\infty}=w\bm{e_1}$, and the corresponding impact factor is denoted by $b~$, as shown in Fig.~\ref{figure1}.

The corresponding Lagrange function in the weak-field limit can be written as
\begin{eqnarray}
\nonumber&&L=\left[1+2(1+v^2)\gamma^2\Phi+(1+\gamma^2)\Phi^2-\frac{4av\gamma^2X_2\Phi}{R^2}
-\frac{v^2\gamma^2X_1^2\Phi^2}{R^2}\right]\!\overset{.}t^{2}  \\
\nonumber&&~~~~~~-\left[(1-\Phi)^2-v^2\gamma^2(4\Phi+\Phi^2)+\frac{4av\gamma^2X_2\Phi}{R^2}
+\frac{\gamma^2X_1^2\Phi^2}{R^2}\right]\!\overset{.}x^{2}\\
\nonumber&&~~~~~~-\left[\left(1-\Phi\right)^2\!+\!\frac{X_2^2\Phi^2}{R^2}\right]\!\overset{.}y^{2}
\!-\!2\!\left[v\gamma^2(4\Phi\!+\!\Phi^2)\!-\!\frac{v\gamma^2X_1^2\Phi^2}{R^2}
\!-\!2a(1+v^2)\gamma^2\frac{X_2\Phi}{R^2}\right]\!\overset{.}t\overset{.}x   \\
&&~~~~~~-2\left(\frac{2a\gamma X_1\Phi}{R^2}-\frac{v\gamma X_1X_2\Phi^2}{R^2}\right)\!\overset{.}t\overset{.}y
-2\left(\frac{\gamma X_1X_2\Phi^2}{R^2}-\frac{2av\gamma X_1\Phi}{R^2}\right)\!\overset{.}x\overset{.}y~,
\end{eqnarray}
where the dot denotes the derivative with respect to $\xi$. After tedious calculations, we can obtain the Lagrange equations up to the second order as follows:
{\small
\begin{align}
\nonumber&2\overset{..}t\!\left[\!1\!+\!2(1\!+\!2v^2\gamma^2)\Phi\!+\!(1\!+\!\gamma^2)\Phi^2\!-\!4av\gamma^2\frac{X_2\Phi}{R^2} \!-\! v^2\gamma^2\frac{X_1^2\Phi^2}{R^2}\!\right]
\!-\!2\overset{..}y\!\left(\!2a\gamma\frac{X_1\!\Phi}{R^2}\!-\!v\gamma\frac{X_1X_2\Phi^2}{R^2}\right)  \\
\nonumber&\!-\!\!2\overset{..}x\!\left[v\gamma^2\!(4\Phi\!+\!\Phi^2)\!-\!v\gamma^2\!\frac{X_1^2\Phi^2}{R^2}
\!-\!2a(1\!+\!2v^2\gamma^2)\frac{X_2\Phi}{R^2}\!\right]\!
\!+\!4\gamma^2\!\!\left[\overset{.}t(1\!+\!v^2)\!-\!2v\overset{.}x\right]\!\!\frac{\partial \Phi}{\partial \xi}
\!-\!4a\gamma\overset{.}y\frac{\partial }{\partial \xi}\!\frac{X_1\!\Phi}{R^2}  \\
\nonumber&\!+\!\!2\!\left[(1\!\!+\!\!\gamma^2)\overset{.}t\!-\!v\gamma^2\overset{.}x\right]\!\!\frac{\partial \Phi^{\!2}}{\partial \xi}
\!+\!4a\gamma^2\!\!\left[(1\!\!+\!v^2)\overset{.}x\!-\!2v\overset{.}t\right]\!\!\frac{\partial }{\partial \xi}\!\frac{X_2\Phi}{R^2}
\!+\!2v\gamma\overset{.}X_{\!1}\frac{\partial }{\partial \xi}\!\frac{X_1^2\Phi^2}{R^2}\!
+\!2v\gamma\overset{.}y\frac{\partial }{\partial \xi}\!\frac{X_1\!X_2\Phi^2}{R^2} \\
\nonumber&\!-\!2\gamma^2\!\!\left[(1\!\!+\!\!v^2)\overset{.}t^2
\!\!+\!2(v^2\overset{.}x^2\!\!-\!2v\overset{.}t\overset{.}x)\right]\!\frac{\partial \Phi}{\partial t}
\!-\!\left[(1\!+\!\gamma^2)\overset{.}t^2\!+\!\gamma^2(v^2\overset{.}x^2\!-\!2v\overset{.}t\overset{.}x)\right]
\!\!\frac{\partial \Phi^2}{\partial t}\!+\!2\overset{.}X_1\overset{.}y\frac{\partial }{\partial t}\!\frac{X_1\!X_2\Phi^2}{R^2}  \\
\nonumber&+(\overset{.}x^2\!+\!\overset{.}y^2)\frac{\partial }{\partial t}(1\!-\!\Phi)^2
\!-\!4a\gamma^2\!\left[(1\!+\!v^2)\overset{.}t\overset{.}x\!-\!v(\overset{.}x^2\!+\!\overset{.}t^2)\right]
\frac{\partial }{\partial t}\!\frac{X_2\Phi}{R^2}\!+\!\overset{.}y^2\frac{\partial }{\partial t}\!\frac{X_2^2\Phi^2}{R^2}
\!+\!4a\overset{.}X_0\overset{.}y\frac{\partial }{\partial t}\!\frac{X_1\Phi}{R^2}  \\
&+\overset{.}X_1^2\frac{\partial }{\partial t}\!\frac{X_1^2\Phi^2}{R^2}=0~, \label{L1}
\end{align}
\begin{align}
\nonumber&-\!2\overset{..}x\!\left[(1\!-\!\Phi)^2\!-\!v^2\gamma^2(4\Phi+\Phi^2)
+4av\gamma^2\frac{X_2\Phi}{R^2}+\gamma^2\frac{X_1^2\Phi^2}{R^2}\right]
\!-\!2\overset{..}y\left(\!\gamma\frac{X_1X_2\Phi^2}{R^2}-2av\gamma\frac{X_1\Phi}{R^2}\!\right)   \\
\nonumber& -2\overset{..}t\!\left[v\gamma^2(4\Phi\!+\!\Phi^2)\!-\!v\gamma^2\frac{X_1^2\Phi^2}{R^2}\!-\!2a(1+v^2)\gamma^2\frac{X_2\Phi}{R^2}\right]
-2\overset{.}x\frac{\partial }{\partial \xi}(1-\Phi)^2-2\gamma\overset{.}y\frac{\partial }{\partial \xi}\frac{X_1X_2\Phi^2}{R^2}  \\
\nonumber&+4a\gamma^2\!\left[(1\!+\!v^2)\overset{.}t\!-\!2v\overset{.}x\right]\frac{\partial }{\partial \xi}\frac{X_2\Phi}{R^2}
\!-\!2\gamma\overset{.}X_1\frac{\partial }{\partial \xi}\frac{X_1^2\Phi^2}{R^2}+4av\gamma\overset{.}y\frac{\partial }{\partial \xi}\frac{X_1\Phi}{R^2}
\!-\!2v\gamma\overset{.}X_0\frac{\partial }{\partial \xi}(4\Phi+\Phi^2)   \\
\nonumber&-2\gamma^2\!\left[(1+v^2)\overset{.}t^2\!+\!2v(v\overset{.}x^2\!-\!2\overset{.}t\overset{.}x)\right]
\!\frac{\partial \Phi}{\partial x}
\!-\!\left[(1+\gamma^2)\overset{.}t^2\!+\!v\gamma^2(v\overset{.}x^2-2\overset{.}t\overset{.}x)\right]\!\frac{\partial \Phi^2}{\partial x}
\!+\!\overset{.}X_1^2\frac{\partial }{\partial x}\frac{X_1^2\Phi^2}{R^2}    \\
\nonumber&-\!4a\gamma^2\!\!\left[\!(1\!+\!v^2)\overset{.}t\overset{.}x\!-\!v(\overset{.}t^2\!\!+\!\overset{.}x^2)\!\right]\!\frac{\partial }{\partial x}\frac{X_2\Phi}{R^2}\!+\!2\overset{.}X_1\overset{.}y\frac{\partial }{\partial x}\frac{X_1\!X_2\Phi^2}{R^2}
\!+\!\overset{.}y^2\frac{\partial }{\partial x}\frac{X_2^2\Phi^2}{R^2}\!+\!4a\overset{.}X_0\overset{.}y\frac{\partial }{\partial x}\frac{X_1\Phi}{R^2}  \\
&+(\overset{.}x^2+\overset{.}y^2)\frac{\partial }{\partial x}(1-\Phi)^2=0~, \label{L2}
\end{align}
\begin{align}
\nonumber&-\!2\overset{..}y\,(1\!-\!2\Phi)+4\overset{.}y\overset{.}\Phi-4a\overset{.}X_0\,\frac{\overset{.}X_1\Phi+3X_1\overset{.}\Phi}{R^2}
\!-\!2\overset{.}X_1\frac{\overset{.}X_1X_2\Phi^2\!+\!4X_1X_2\overset{.}\Phi\Phi}{R^2}+\overset{.}X_1^2X_1^2\,\frac{\partial}{\partial y}\frac{\Phi^2}{R^2}     \\
\nonumber&-\!2\left[\left(1\!+\!2v^2\gamma^2\right)\left(\overset{.}t^2\!+\!\overset{.}x^2\right)
-4v\gamma^2\overset{.}t\overset{.}x\,\right]\frac{\partial \Phi}{\partial y}
\!-\!\left[(1+\gamma^2)\,\overset{.}t^2+(v^2\gamma^2-1)\,\overset{.}x^2-2v\gamma^2\overset{.}t\overset{.}x\right]\frac{\partial \Phi^2}{\partial y}   \\
&-4a\gamma^2\left[(1+v^2)\overset{.}t\overset{.}x-v\left(\overset{.}t^2+\overset{.}x^2\right)\right]
\frac{\partial}{\partial y}\frac{X_2\Phi}{R^2}=0~. \hspace*{30pt} \label{L3}
\end{align}}
Here $X_i~(i=0,1,2)$ is defined in Eq.~\eqref{LorentzTran1}. The affine parameter $\xi=x$ has been used in Eq.~\eqref{L3}.

Since the high-order deflection by Schwarzschild or the non-moving Kerr black hole have been studied in detail, here we primarily investigate the leading roto-translational effects on the deflection for our focus. So the bending angle to be derived doesn't stand for total deflection angle up to second order now that we don't consider pure second-order Schwarzschild terms. Under this circumstance, we notice the conditions to be used still keeps its form like
that in Ref.~\cite{Olaf2004} as follows:
\begin{eqnarray}
&& dx=\left[\frac{1}{\gamma\left(1-\frac{v}{w}\right)}+O(m)\right]dX_1~,\label{condition1} \\
&&\overset{.}t=\frac{1}{w}+O(m)~,\label{condition2}  \\
&&\overset{.}x=1+O(m^2)~. \label{x}
\end{eqnarray}
Here, $\xi$ has been replaced by $x$ to derive them from Eqs.~\eqref{L1}\,-\,\eqref{L2}.

So the corresponding motion equation can be obtained from Eqs. \eqref{L3}\,-\,\eqref{x} as
\begin{eqnarray}
\nonumber&&\overset{..}y=\!-\!\left[(1+v^2)(1+\frac{1}{w^2})-\frac{4v}{w}\right]\!\gamma^2\Phi_{,X_2}
\!-\!2a\gamma^2\!\left[\frac{1+v^2}{w}-v\left(1\!+\!\frac{1}{w^2}\right)\right]\frac{\Phi\!+\!3X_2\Phi_{,X_2}}{R^2} \\
&&~~~~\,+2a\gamma^2\left(v-\frac{1}{w}\right)\frac{\Phi+3X_1\Phi_{,X_1}}{R^2}~, ~\label{jiasudu}
\end{eqnarray}
where $\Phi_{,X_i}$ denotes the partial differential $\frac{\partial \Phi}{\partial X_i}$. We assume velocity $v$ of the uniformly moving Kerr black hole is less than $w$. Therefore, the deflection angle of the test particle with kinematic corrections can be obtained via integrating from $-\infty$ to $+\infty$:
{\small
\begin{eqnarray}
\!\alpha(a,v,w)\!=\!N_1\!(v,w)\!\!\left[-(1\!+\!\frac{1}{w^2})\!\!\int^{+\infty}_{\!-\infty}\!\!\Phi_{,X_2}\,d X_1\!\right]
\!+\!N_2(v,w)\!\!\left[-\frac{2a}{w}\!\!\int^{+\infty}_{\!-\infty}\!\frac{\Phi\!+\!3X_2\Phi_{,X_2}}{R^2}d X_1\!\right]\!,~~~\label{angle1}
\end{eqnarray}}
where  $N_1(v,w)~$ and $~N_2(v,w)$ represent the coefficients of kinematic corrections:
\begin{eqnarray}
&&N_1(v,w)=\frac{\gamma\left(1+v^2-\frac{4vw}{1+w^2}\right)}{1-\frac{v}{w}}~, \label{coefficentN1} \\
&&N_2(v,w)=\gamma\left(1-wv\right)~.   \label{coefficentN2}
\end{eqnarray}

We finally obtain the explicit expression for Eq.~\eqref{angle1} as follow
\begin{eqnarray}
\alpha (a,v,w)=N_1(v,w)\left[\left(1+\frac{1}{w^2}\right)\frac{2m}{b}\right]+N_2(v,w)\left[-\frac{1}{w}\frac{4ma}{b^2}\right]~,~~~\label{angle2}
\end{eqnarray}
where we have made use of the relation $|X_2|=-X_2\approx\sqrt{X_2^2-a^2}\approx b$ in the limit of small-angle approximation.

For the light propagating in the field of a non-moving Kerr black hole, we have $v=0$ and Eq.~\eqref{angle2} is reduced to the well-known result~\cite{BHCQR2004,BHPRD2004,BCQR2005,SerenoLuca2006,AazKeePett2011}
\begin{eqnarray}
\alpha(a,0,1) = \frac{4m}{b}-\frac{4ma}{b^2}~,\label{zuihoujieguo}
\end{eqnarray}
The contribution of rotating term in Eq.~\eqref{zuihoujieguo} is negative when the photon takes the prograde motion relative to the rotation of Kerr black hole, as shown in Fig.~\ref{figure1}. The contribution will become positive when the photon takes the retrograde motion relative to the rotation $(a<0)$.

\begin{figure}
\setlength{\unitlength}{1cm}
\begin{minipage}[b]{12.5cm}
  \centering
  \includegraphics[width=10.5cm]{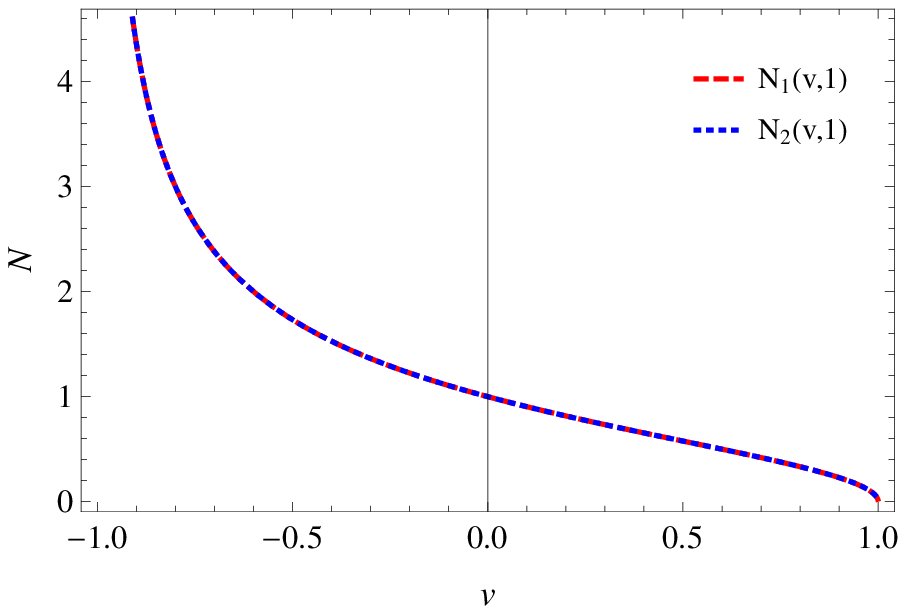}
  \centerline{$(a)$}
\end{minipage}
\begin{minipage}[b]{12.5cm}
  \centering
  \includegraphics[width=11.2cm]{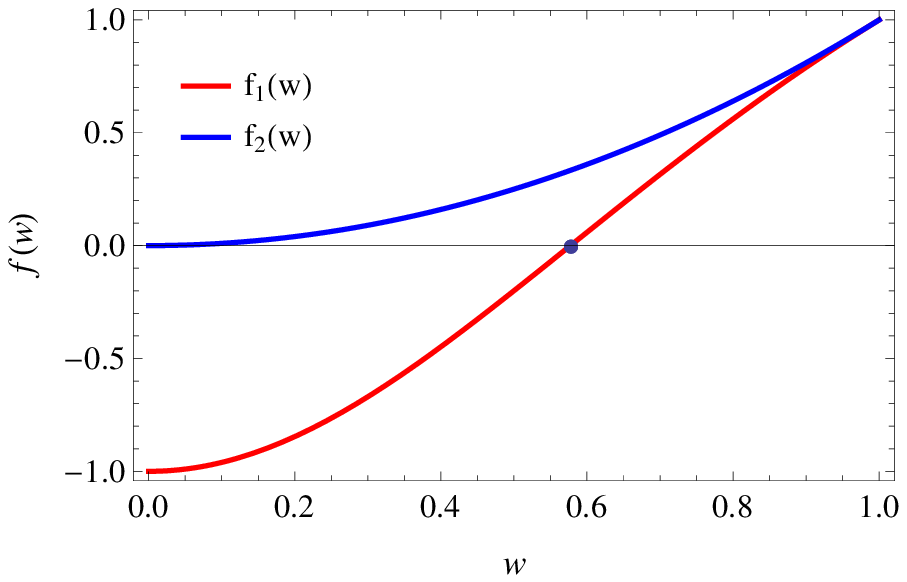}
  \centerline{$(b)$}
\end{minipage}

\caption{Two coefficients $N_1(v,1)$ and $N_2(v,1)$ changing with $v$ (left panel) and the behavior of internal function $~f_2(w)=w^2~$ of the second coefficient $~N_2(v,w)~$ compared with $~f_1(w)=\frac{3w^2-1}{1+w^2}$ (right panel)~.} \label{figure2}
\end{figure}

\section{The leading roto-translational effects} \label{kinematic discussion}
Wucknitz and Sperhake have studied the translational effects on the first-order deflection of the test particle in detail~\cite{Olaf2004}. Here we will investigate the leading compound roto-translational effects, based on Eq.~\eqref{angle2}.

\subsection{Discussions of velocity effects}
For light $(w=1)$, Eq.~\eqref{angle2} is simplified to
\begin{eqnarray}
\alpha(a,v,1)=\gamma(1-v)\frac{4m}{b}-\gamma(1-v)\frac{4ma}{b^2}~.\label{angle3}
\end{eqnarray}
As discussed about scaling factor by Sereno~\cite{Sereno2005b}, $N_2(v,1)$ is the same value as the first-order coefficient $~N_1(v,1)$. So they have the same convergence and divergence, namely diverging for $~v\rightarrow -1~$ while vanishing for $~v\rightarrow +1$, as showed visibly in Fig.~\ref{figure2}. Nevertheless, it's amazing that this homogeneity doesn't apply to any other massive test particle. This difference leads to the privileged position of light among all test particles when one weighs the magnitude of second-order Kerr deflection (SOKD) with correction.

In the limit of small translational velocity (\,$|\frac{v}{w}| \rightarrow 0$\,), namely the first-order velocity (FOV) approximation, $N_2(v,w)$ can be simplified as
\begin{eqnarray}
N_2(v,w) \approx 1-wv~.  \label{N2}
\end{eqnarray}
Following the work of Ref. \cite{Olaf2004}, the internal function $f_i(w)$ is defined as $f_i(w) = \left[1-N_i(v,w)\right]\frac{w}{v}$. The corresponding internal function $~f_2(w)~$ of $~N_2(v,w)~$ is showed in Fig. \ref{figure2}, to compare with the internal function in $~N_1(v,w)$. In the limit of FOV approximation, one can obviously observe that the coupling between translational motion and rotation always persists since $f_2(w)>0$ for all test particles.

Actually, we find this coupling persists all the time and it doesn't limit to special circumstances such as FOV approximation because both solutions of equation below
\begin{equation}
N_2(v,w)=1~,
\end{equation}
don't satisfy our basic physical conditions of $v<w,~-1\leq v<1$ and $0<w$. Therefore, we argue that the effects of translational velocity always exists in leading rotational term and may also always appear in total deflection angle up to second order, for any test particle including light.

Eq.~\eqref{angle2} for the limit of FOV approximation is reduced to
\begin{eqnarray}
\alpha(a,v,w)_{FOV}=\left(1+\frac{1}{w^2}\right)\frac{2m}{b}-\frac{1}{w}\frac{4ma}{b^2}
+\frac{2v(1-3w^2)}{w^3}\frac{m}{b}+\frac{4vma}{b^2}~,~\label{angle4}
\end{eqnarray}
which includes four terms on the right side and is consistent with Eq.~(27) in Ref.~\cite{Olaf2004}. We emphasize that formula Eq.~\eqref{angle4} here applies to test particles with velocity $w$ which satisfies the prerequisite $~|\frac{v}{w}|\ll 1~$ and small angle approximation. The first two terms on the right side of Eq.~\eqref{angle4} are basic first-order Schwarzschild deflection and SOKD for test particle. The third term arises from pure translational effects of uniformly moving Kerr black hole. The rest is the coupling term between translational motion and rotation, which persists invariably as long as $v\neq 0$. For light it takes the simple form of
\begin{eqnarray}
\alpha(a,v,w)_{FOV}=(1-v)\left(\frac{4m}{b}-\frac{4ma}{b^2}\right)~,  \label{angle5}
\end{eqnarray}
which can be also obtained from Eq.~\eqref{angle3} in the limit of FOV approximation and is consistent with previous works (see, e.g., Refs. \cite{PyneBirk1993,Frittelli2003}).
\begin{figure}
\setlength{\unitlength}{1cm}
\begin{minipage}{12.5cm}
  \centering
  \includegraphics[width=10cm]{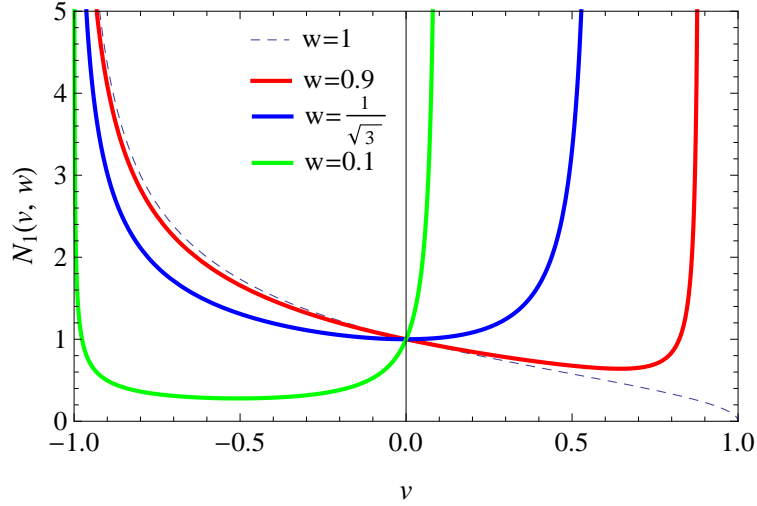}
  \centerline{$(a)~N_1(v,w)$}     \label{figure3-1}
\end{minipage}
\begin{minipage}{12.5cm}
  \centering
  \includegraphics[width=10cm]{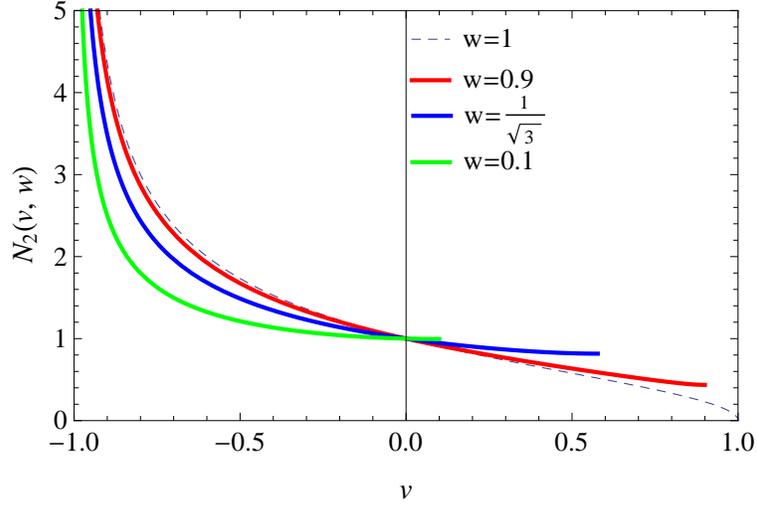}
  \centerline{$(b)~N_2(v,w)$}     \label{figure3-2}
\end{minipage}

\caption{The general behavior of coefficient $~N_2(v,w)~$ as function of two parameters $v$ and $w$. We still choose four typical values $~0.1,~1/\sqrt{3}~(\approx 0.577),~0.9,~$ and $~1~$ for velocity $~w~$ of test particle in order to compare with $~N_1(v,w)$.} \label{figure3}
\end{figure}

With respect to $~w=1/\sqrt{3}$, although pure effects of translational motion of Kerr black hole in the first term of Eq.~\eqref{angle2} vanish for FOV approximation, $N_2(v,w)$ depends on translational velocity $~v$. Thus, for $~w=1/\sqrt{3}~~$(FOV), all of the pure effects of rotation, pure effects of velocity and the coupling effects mentioned above may still persist in total second-order deflection angle (including pure second-order Schwarzschild terms) besides the basic term in Eq.~\eqref{angle4}, which is totally different from the case of first-order deflection.

In the limit of $v=0$ and $w>0~$, deflection angle for test particle including light by Kerr black hole from Eq.~\eqref{angle2} is
\begin{eqnarray}
\alpha(a,0,w)=\left(1+\frac{1}{w^2}\right)\frac{2m}{b}-\frac{1}{w}\frac{4ma}{b^2} ~,    \label{angle6}
\end{eqnarray}
which can be found directly in Eq.~\eqref{angle4} and recovers Eq.~\eqref{zuihoujieguo} for light. What it predicts at once is an evident conclusion that rotational term is inversely proportional to test particles' velocity, namely the absolute contribution of rotational term of light takes minimum value among all tests of deflection of test particles. While the mentioned black hole moves in a clockwise direction $(a<0)~$ in Fig.~\ref{figure1}, deflection angle increases with decreasing velocity $w$ of test particle. Therefore, one may use the massive particles cast by massive celestial body to instead of photon in gravitational lensing in some cases if possible. For example, if one sets $w=0.1~$, the first-order deflection angle will increase by about 100 times and contribution of second-order deflection enlarges 10 times. The advantage is obvious for measuring total bending angle in the area of astrophysics.

For general cases of $v<w$, the graphic of $~N_2(v,w)~$ is plotted in Fig.~\ref{figure3} compared with ratio $~N_1(v,w)$. It is obvious that $~N_2(v,w)~$ diverges for the limit $v\rightarrow -1$ while takes finite value for the limit $v\rightarrow w$.

\subsection{The perturbation of roto-translational effects to the contribution of SOKD} \label{rotational effect}
\begin{figure}[b]
\setlength{\unitlength}{1cm}
\begin{minipage}[b]{12.5cm}
  \centering
  \includegraphics[width=9.5cm]{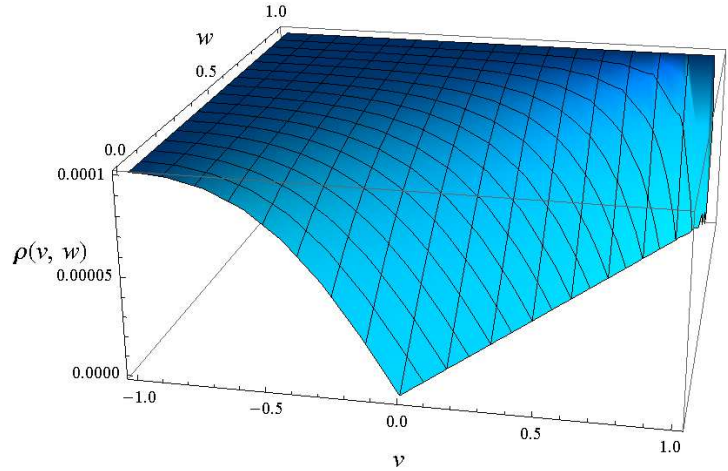}
  \centerline{$(a)~\chi=10^{-4}$}     \label{figure4-1}
\end{minipage} \\
\begin{minipage}[b]{12.5cm}
  \centering
  \includegraphics[width=9.5cm]{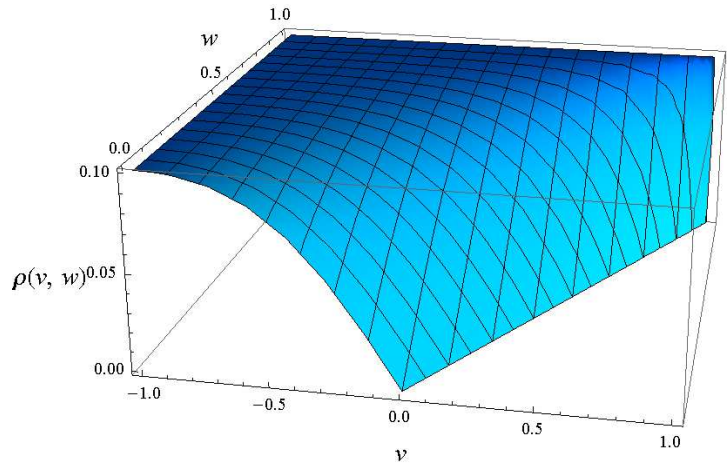}
  \centerline{$(b)~\chi=10^{-1}$}     \label{figure4-2}
\end{minipage}
\caption{$\rho$ as the function of two variables $v$ and $w~$. Here the basic assumption $~v<w~$ is adopted. We choose $10^{-4}$ and $0.1$ for $\chi$ as examples.} \label{figure4}
\end{figure}
\begin{figure}
\setlength{\unitlength}{1cm}
\begin{minipage}[b]{6.2cm}
  \centering
  \includegraphics[width=6.2cm]{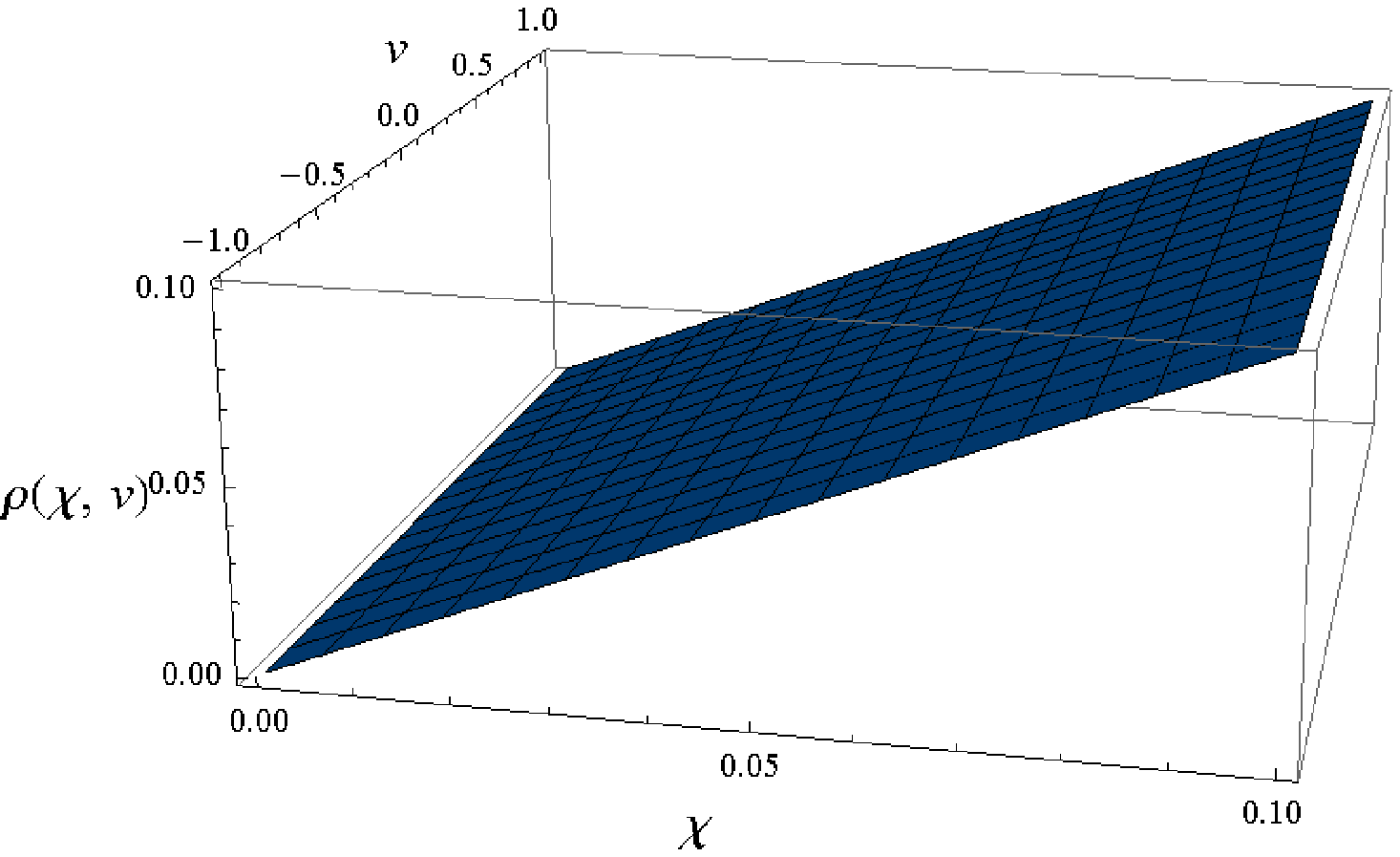}
  \centerline{$(a)~w=1$}     \label{figure5-1}
\end{minipage}
\begin{minipage}[b]{6.2cm}
  \centering
  \includegraphics[width=6.2cm]{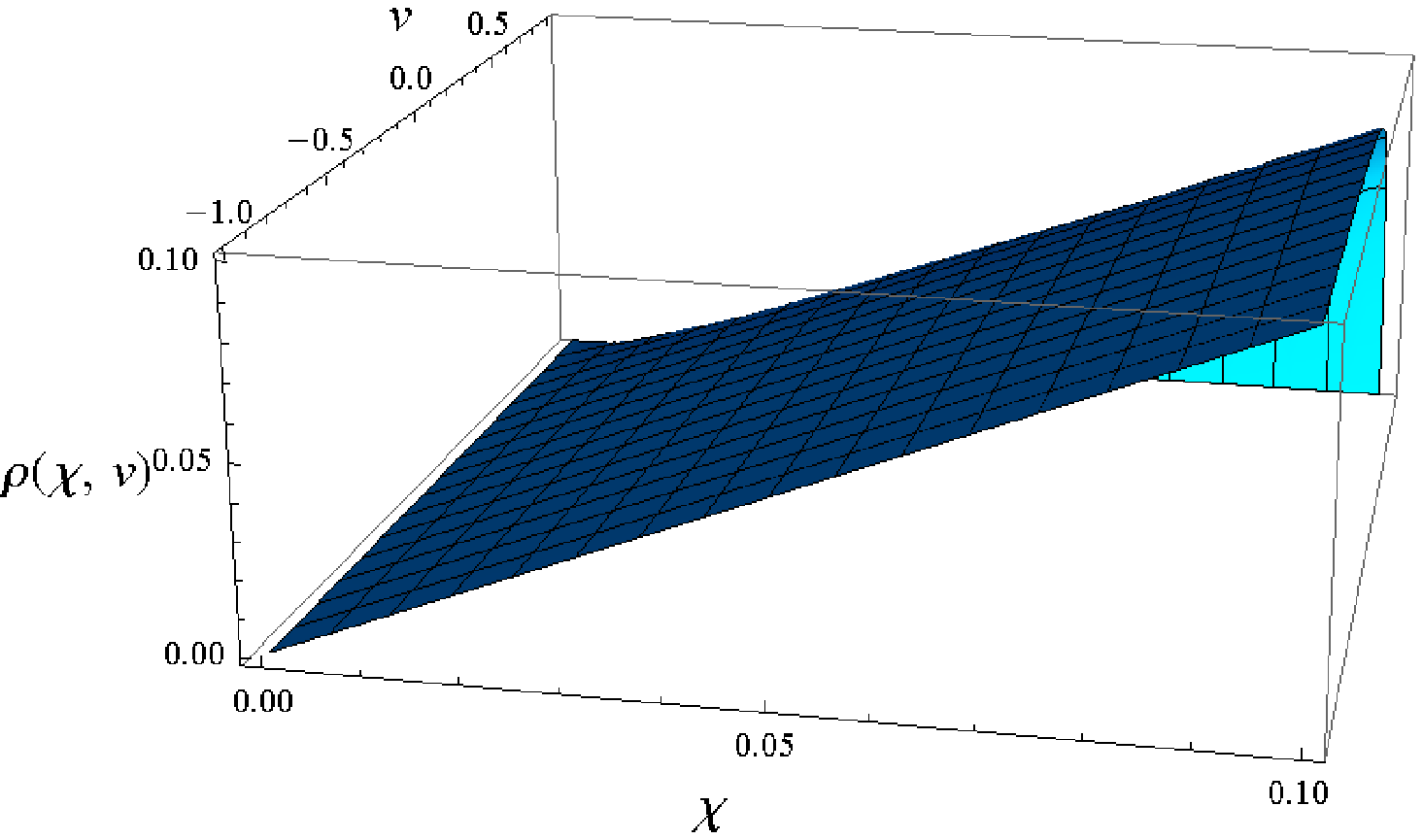}
  \centerline{$(b)~w=0.9$}     \label{figure5-2}
\end{minipage}  \\
\begin{minipage}[b]{6.2cm}
  \centering
  \includegraphics[width=6.2cm]{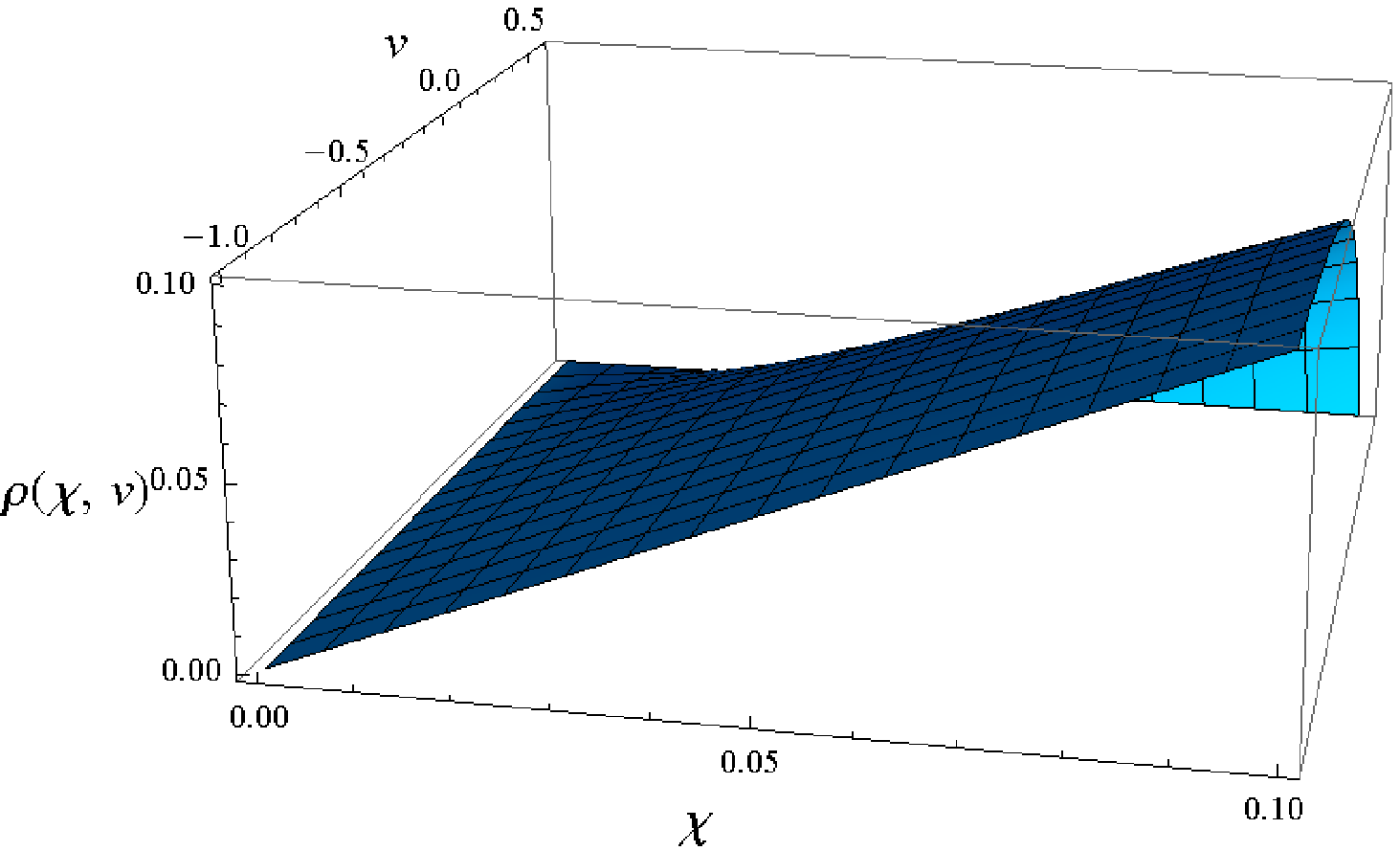}
  \centerline{$(c)~w=1/\sqrt{3}$}   \label{figure5-3}
\end{minipage}
\begin{minipage}[b]{6.2cm}
  \centering
  \includegraphics[width=6.2cm]{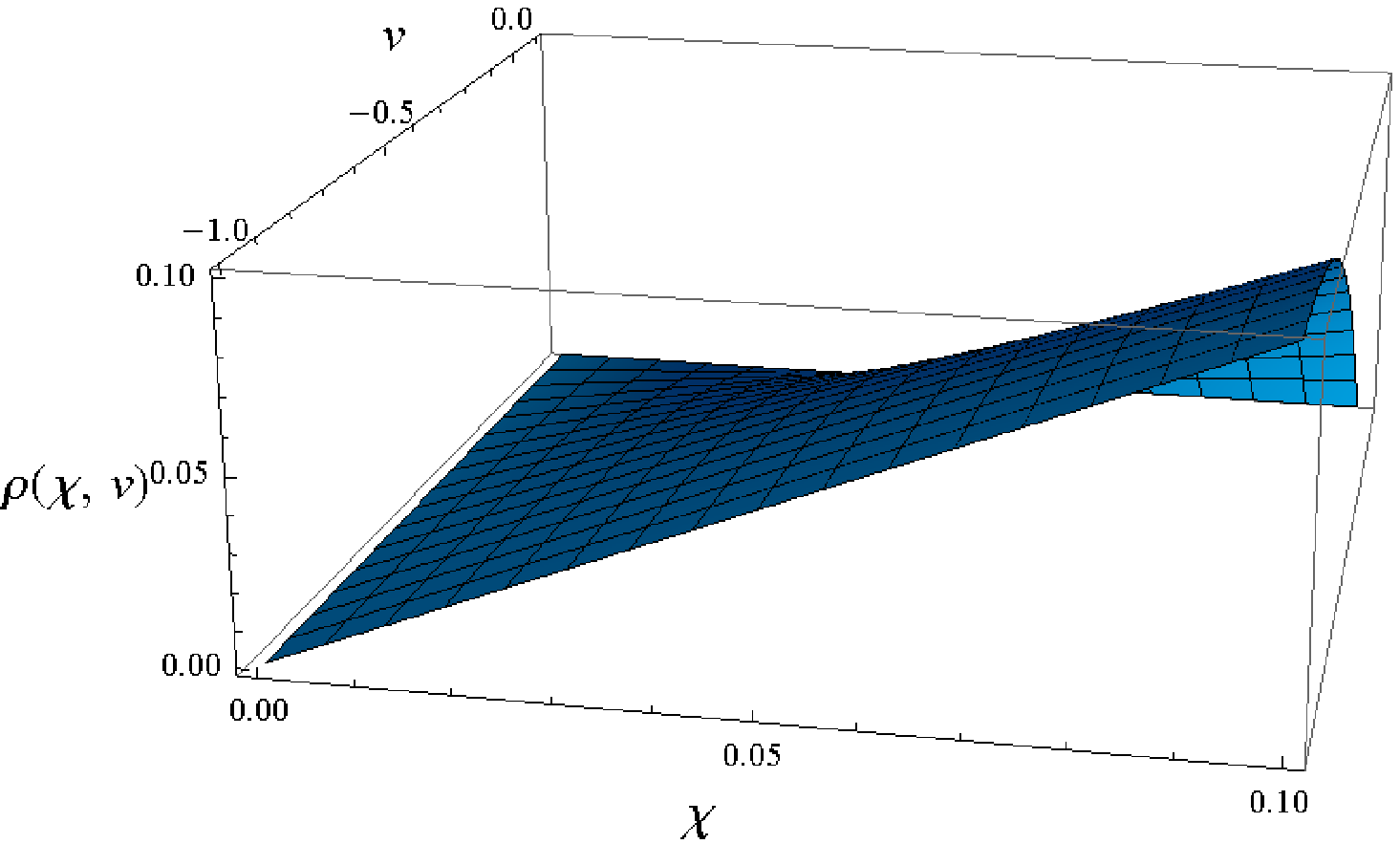}
  \centerline{$(d)~w=0.1$}     \label{figure5-4}
\end{minipage} \\
\begin{minipage}[b]{6.2cm}
  \centering
  \includegraphics[width=6.2cm]{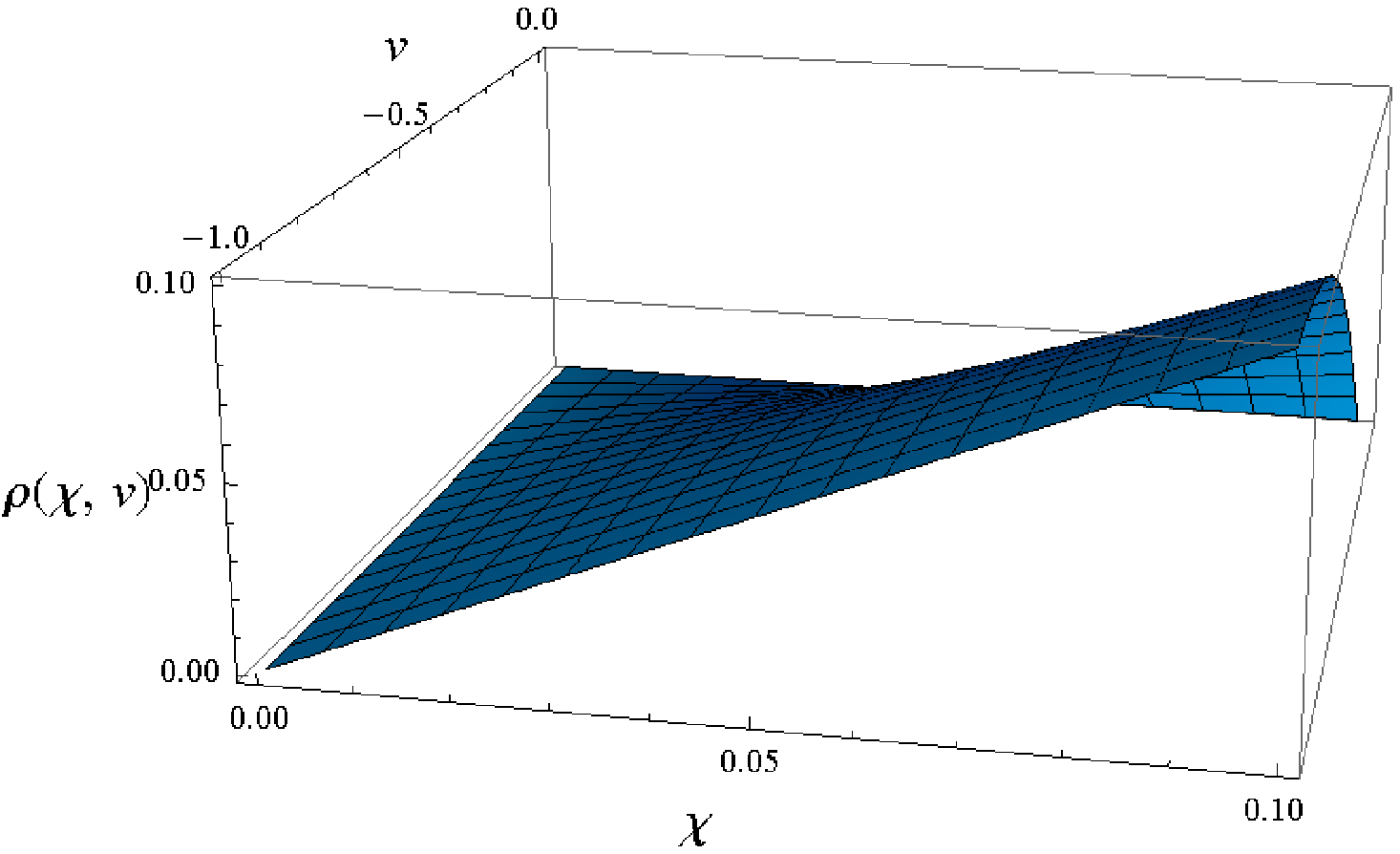}
  \centerline{$(e)~w=0.01$}     \label{figure5-5}
\end{minipage}
\caption{General behavior of $\rho$ as the function of two variables $\chi$ and $v\,(<w)$. $~w~$ takes five typical values of $~1,~0.9,~0.577,~0.1$ and $~0.01$. As an example, here we set $\chi\leq 0.1$ which is also used in Fig. \ref{figure6}.} \label{figure5}
\end{figure}
\begin{figure}
\setlength{\unitlength}{1cm}
\begin{minipage}[b]{6.2cm}
  \centering
  \includegraphics[width=6.2cm]{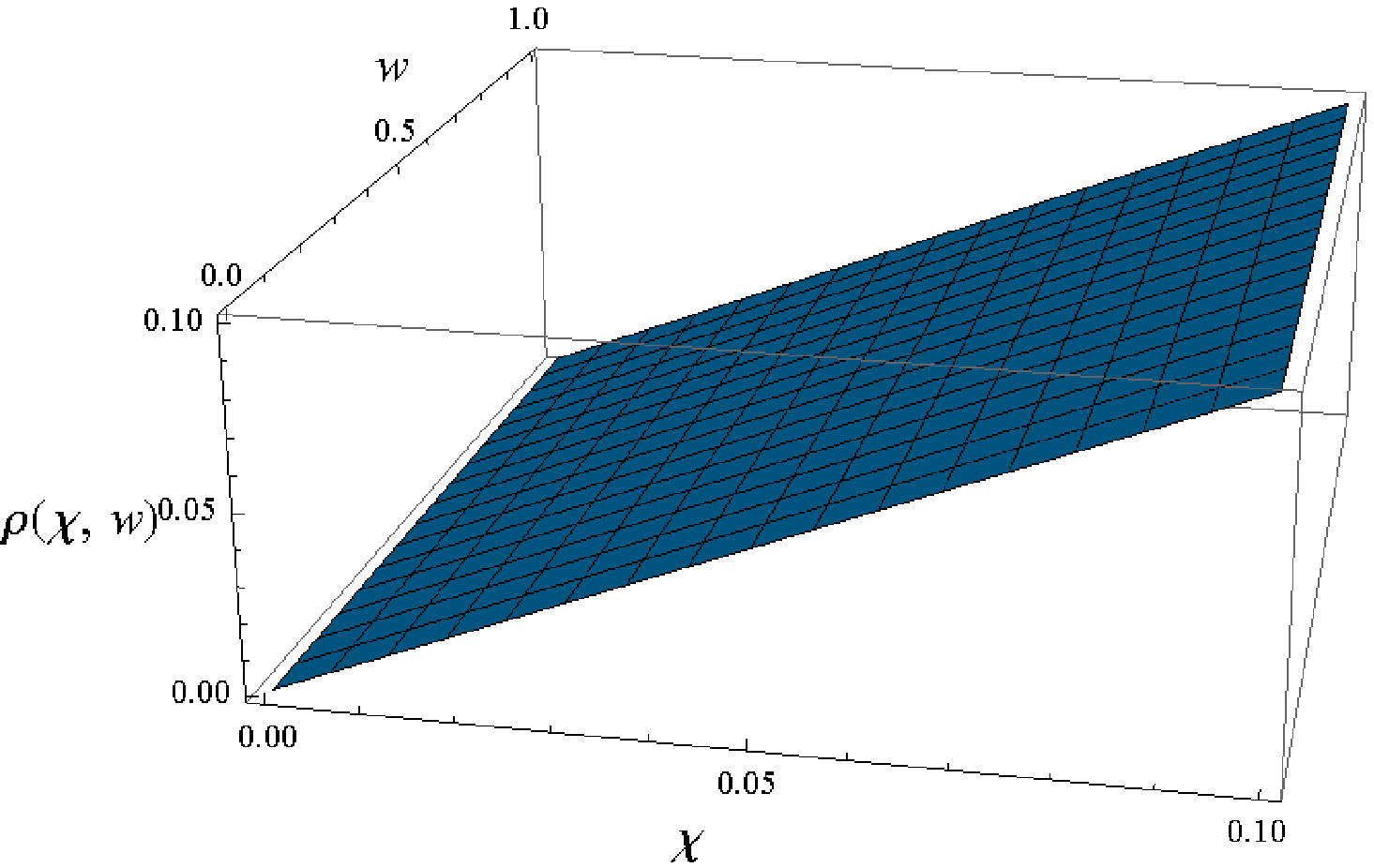}
  \centerline{$(a)~v=-0.99$}     \label{figure6-1}
\end{minipage}
\begin{minipage}[b]{6.2cm}
  \centering
  \includegraphics[width=6.2cm]{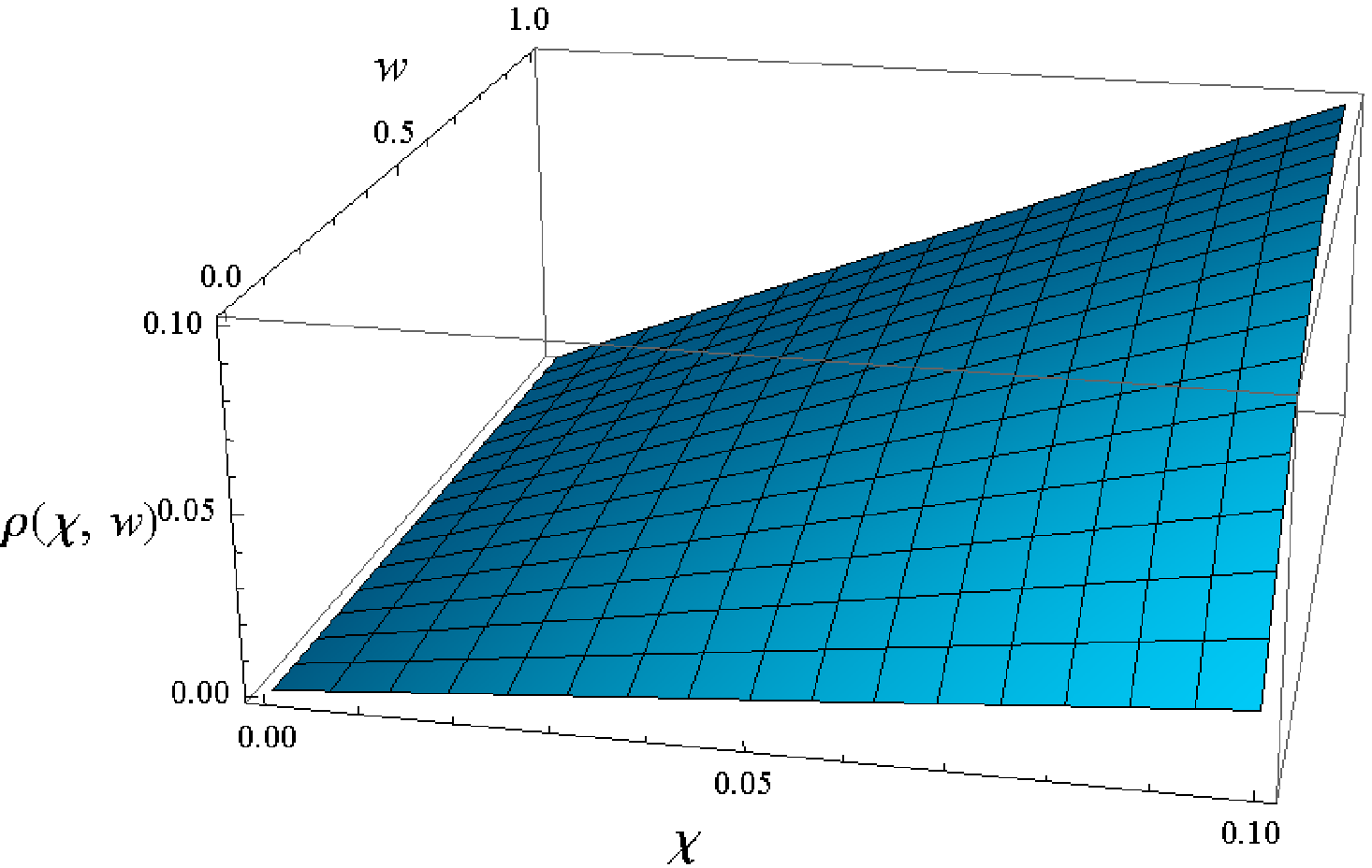}
  \centerline{$(b)~v=-0.1$}   \label{figure6-2}
\end{minipage}  \\
\begin{minipage}[b]{6.2cm}
  \centering
  \includegraphics[width=6.2cm]{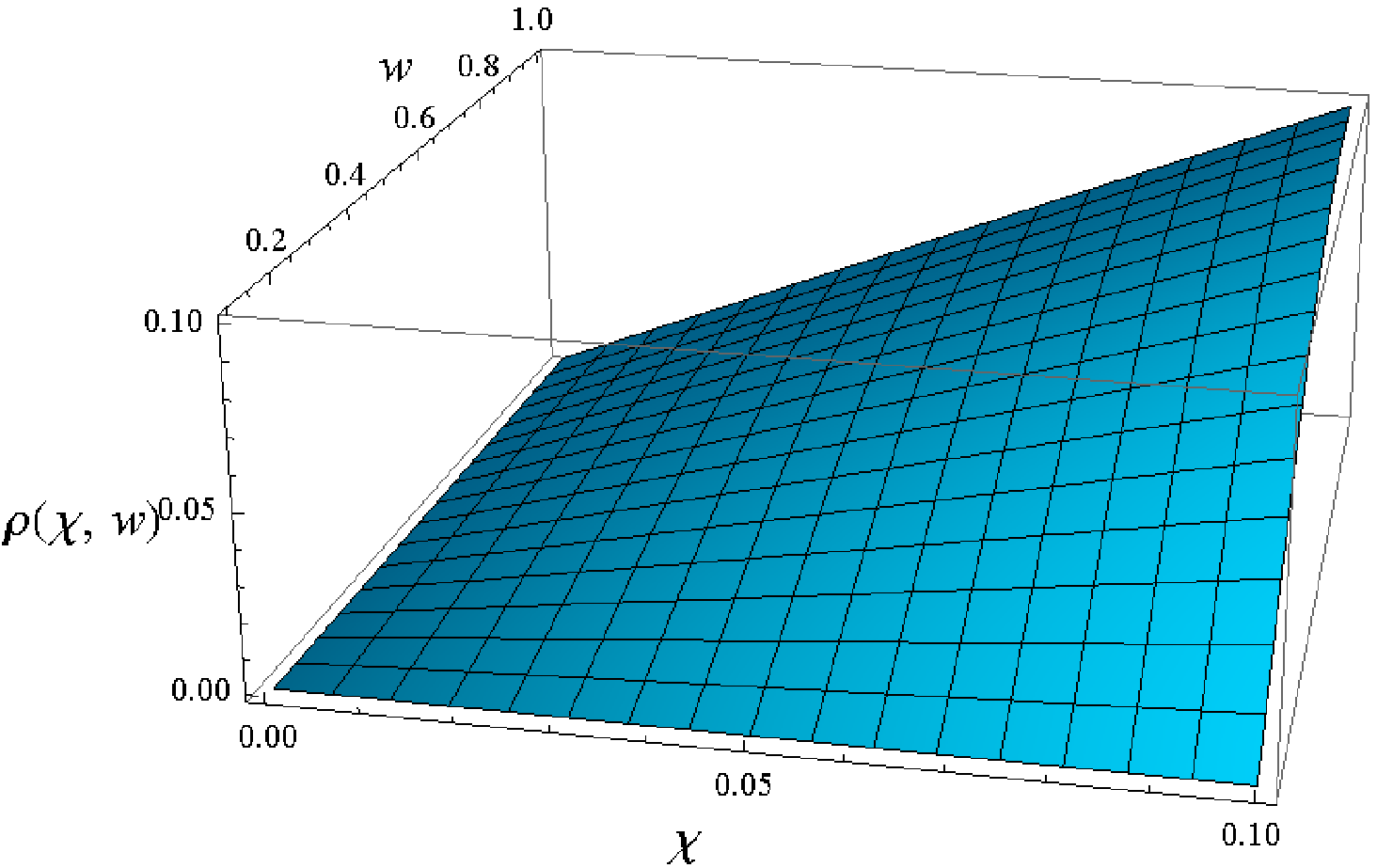}
  \centerline{$(c)~v=0.1$}   \label{figure6-3}
\end{minipage}
\begin{minipage}[b]{6.2cm}
  \centering
  \includegraphics[width=6.2cm]{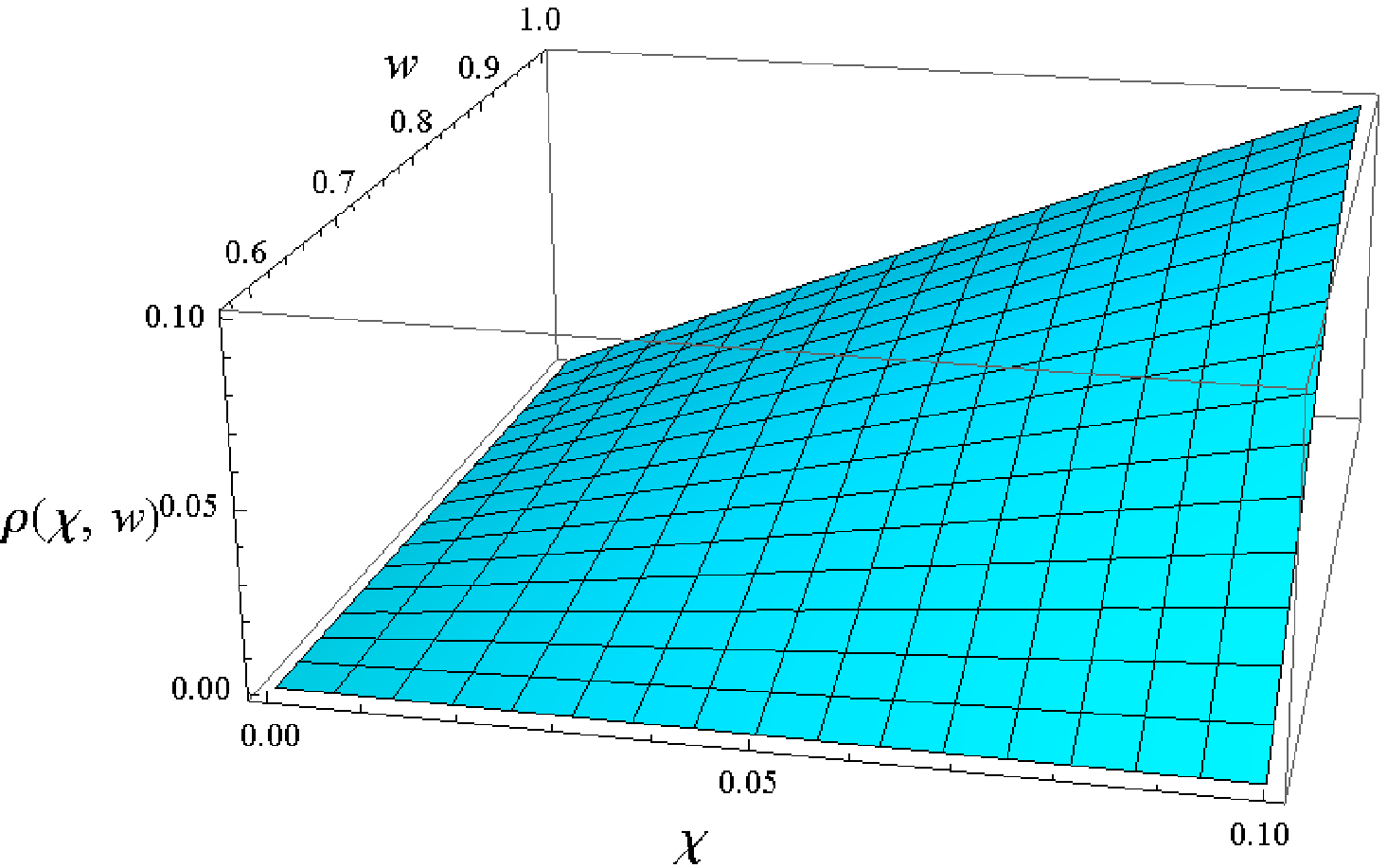}
  \centerline{$(d)~v=0.577$}     \label{figure6-4}
\end{minipage} \\
\begin{minipage}[b]{6.2cm}
  \centering
  \includegraphics[width=6.2cm]{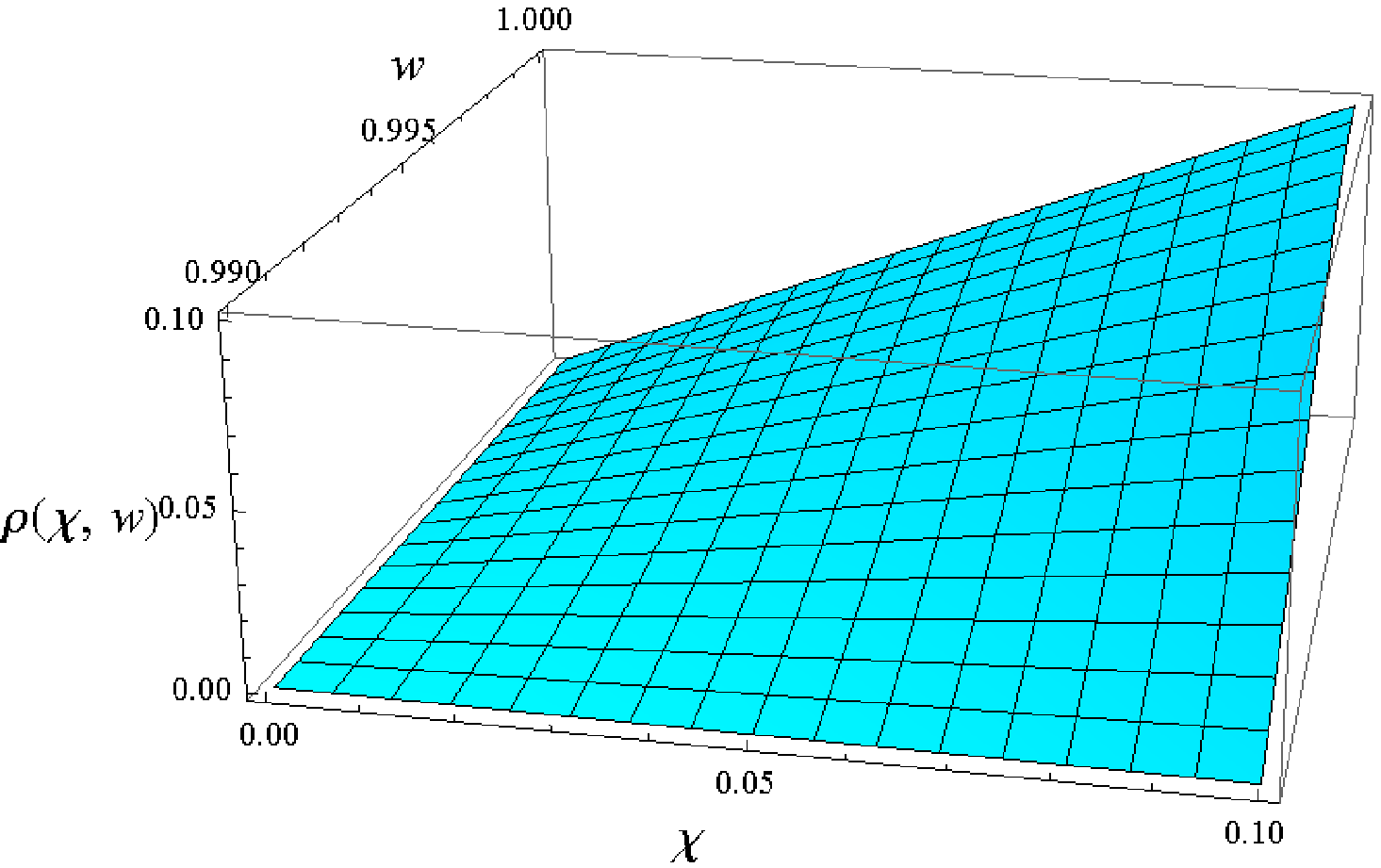}
  \centerline{$(e)~v=0.99$}     \label{figure6-5}
\end{minipage}
\caption{$\rho$ as the function of two variables $\chi$ and $w(>v)$. Here we preset five values, namely $-0.99,\,-0.1,~0.1,~0.577$ and $~0.99~$ for $v$.} \label{figure6}
\end{figure}
In this section, we calculate the contribution of the leading rotational terms to the deflection angle, which can be characterized by the ratio of the kinematically corrected SOKD relative to the first-order deflection in Eq.~\eqref{angle2} as
\begin{eqnarray}
\rho(\chi,v,w)=\frac{2\left[1+v^2-v\left(w+\frac{1}{w}\right)\right]}{\left(1+v^2-\frac{4vw}{1+w^2}\right)\left(w+\frac{1}{w}\right)}\chi~,  \label{N1N6}
\end{eqnarray}
where $\chi=a/b$ denotes a function of weighing rotational parameter $a$ for some impact factor. Here, we discuss the case of $a>0$ without loss of generality, so that $\rho$ is non-negative. With the help of $\rho$, it's convenient to grope for the corrections of roto-translational effects to the contribution mentioned above. A combination of variables like $(\chi,v,w)$ is used to clarify the dependence.

For a moving Kerr black hole with certain $\chi$, the perturbation of translational effects to the mentioned contribution is approached the upper bound asymptotically and no peak emerges, as showed in Fig. \ref{figure4}. Concretely, we find the parameter $\rho$ takes a constant and maximum value $\chi$ for light, while it strictly decreases monotonically with increasing $v$~for any other test particle. In other words, the contribution of kinematically corrected SOKD will become larger for a certain test particle, if black hole moves with a smaller speed toward positive x-axis. $\rho$ approaches the upper bound $\chi$ when Kerr black hole moves along negative $x$-axis with a highly relativistic velocity ($v\rightarrow -1$). And, $\rho$ will turn to its minimum $0$ in the limit of $v\rightarrow w~(w\neq 1)$. On the other hand, in case the black hole with certain $t$ moves at a given speed $v~(-1<v<w)$, namely for the same black hole, we find $\rho$ will strictly increase monotonically with increasing test particle's velocity $w$. Therefore, provided that one use light as test particle, the mentioned contribution will be most noticeable even if it's still so small in general. Different from that in the limit $v\rightarrow w~(w\neq 1)$, the minimum of $\rho$ is positive for $v<0$. For the limit $v\rightarrow -1$, $\rho$ is independent on $w$~. Hence, there exist three special isolines of $\rho$ in Fig. \ref{figure4}, i.e., the zero line $w=v$ and two maximum lines including the line $w=1$ for the upper limit and the ultrarelativistic line $v=-1$. They restrict the variation range of the contribution of the kinematically corrected SOKD. As regards the shape as a whole due to the perturbation of velocities of both lens and test particle, it looks as if someone dragged one side of the two-dimensional trapezoid down.

Now we switch to investigate the general behaviors of parameter $\rho$ as the function of two variables $\chi$ and $v$ for certain test particle, as performed in Fig.~\ref{figure5}. Similar to the analyses above, $\rho$ for light $(w=1)$ becomes one two-dimensional rectangle when overlooking it. For a given translational velocity $v$, $\rho$ is simplified to an oblique line with certain massive test particle. Compared to the case mentioned above for shape, the side to be dragged belongs to this oblique rectangle.

We also exhibit the characteristics of the parameter $\rho$ from another perspective, i.e. $\rho$ as the function of $\chi$ and $w$, as shown in Fig. \ref{figure6}.

\section{1.5\,PN dynamics for test particles}\label{dynamics}
Based on Eqs. \eqref{g00m}\,-\,\eqref{gijm}, we can derive explicit geodesic equation of a massive particle up to
order $1/R^{3.5}$ as follow
{\small
\begin{align}
\nn&\frac{d\bm{u}}{dt}=\!-\nabla\!\left\{(1\!+\!v^2)\gamma^2\Phi
\!+\!\!\left[\frac{1\!+\!\gamma^2}{2}\!+\!(1\!+\!v^2)\gamma^2\right]\!\Phi^2
\!+\!v\gamma^2\zeta_1\!-\!\frac{1}{2}v^2\gamma^2\frac{X_1^2\Phi^2}{R^2}\right\}\!\!-\!\frac{\partial \bm{\xi}}{\partial t}
\!+\!\bm{u}\!\times\!\left(\nabla\!\times\!\bm{\xi}\right)  \\
\nn&+\!\left[2\!+\!(1\!+\!v^2)\gamma^2\right]\bm{u}\frac{\partial \Phi}{\partial t}\!+\!2\Phi\bm{u}\!\times\!(\nabla\!\times\!\bm{\xi})
+2\left[1\!+\!(1\!+\!v^2)\gamma^2\right]\!\bm{u}(\bm{u}\cdot\nabla)\Phi-(\bm{u}^2\!+\!2v^2\gamma^2u_1^2)\nabla\Phi \\
\nn&-4v\gamma^2u_1\bm{u}\,(\bm{u}\cdot\nabla)\,\Phi+2v(1+v^2)\gamma^4\bm{u}\,\frac{\partial \Phi^2}{\partial x}
+\left\{4v\gamma^2\left[vu_1+(1-2v^2\gamma^2)\Phi\right]\!\left(\frac{\partial}{\partial t}\!+\!\bm{u}\!\cdot\!\nabla\right)\!\Phi \right.\\
&\left. +2v\gamma^2\!\left[(1+v^2)\gamma^2\,\bm{u}\cdot\nabla-v\gamma^2(1+v^2-4vu_1)\frac{\partial}{\partial x}
-2\frac{\partial}{\partial t}\right]\Phi^2\right\}\bm{e_1}~,  \label{d1}
\end{align}}
where $\bm{u}$ stands for three-dimensional velocity vector of test particle and it has been assumed to be order $1/R^{\frac{1}{2}}$, $u_i$ is used to denote $u^i\equiv\frac{dx^i}{dt}$ for convenience and
{\small
\begin{align}
&\bm{\xi}=\!\left[\!v\gamma^2\!\left(\!4\Phi\!+\!\Phi^2\!-\!\frac{X_1^2\Phi^2}{R^2}\right)\!+\!(1\!+\!v^2)\gamma^2\zeta_1\!\right]\!\bm{e_1}
\!+\!\left(\!\gamma\zeta_2\!-v\gamma\frac{X_1X_2\Phi^2}{R^2}\right)\!\bm{e_2}\!-\!v\gamma\frac{X_1X_3\Phi^2}{R^2}\bm{e_3}~,\\
&\zeta_1=\frac{2maX_2}{R^3}~, \\
&\zeta_2=-\frac{2maX_1}{R^3}~.
\end{align}}
In the limit of $v \rightarrow 0~$, $\bm{\xi}$ reduces to
\begin{eqnarray}
\bm{\xi}=\zeta_1\,\bm{e_1}+\zeta_2\,\bm{e_2}~.
\end{eqnarray}

Up to order $1/R^{2.5}$, the corresponding lightlike dynamics is
\begin{align}
\nn&\frac{d\bm{u}}{dt}=-\!\left[\,\bm{u}^2+(1+v^2)\gamma^2+2v^2\gamma^2u_1^2\,\right]\nabla\Phi
+2\left[\,1+(1+v^2-2vu_1)\gamma^2\,\right]\bm{u}\,(\bm{u}\cdot\nabla)\Phi \\
&+\bm{u}\!\times\!\left(\nabla\!\times\!\bm{\xi}\right)+\!\left[2\!+\!(1\!+\!v^2)\gamma^2\right]\bm{u}\frac{\partial \Phi}{\partial t}
+4v\gamma^2\!\left[vu_1\left(\frac{\partial}{\partial t}+\bm{u}\cdot\nabla\right)\Phi
-\frac{\partial \Phi}{\partial t}\right]\bm{e_1}~.\label{d2}
\end{align}
Note that both $\bm{\xi}$ in the expression of $2\Phi\bm{u}\times(\nabla\times\bm{\xi})$ in Eq.~\eqref{d1} and $\bm{\xi}$ in Eq.~\eqref{d2} are simplified to be $4v\gamma^2\Phi\bm{e_1}$.

\section{Applications in astrophysics and astronomy}\label{applications}
Single moving Kerr black hole may be created by a merger in binary systems~\cite{MiloPhi2005,Pretorius2005,Cua2009,Lovelace2010} or gravitational collapse of massive celestial body~\cite{Hawking1974,Hawking1976,Nakamura1981,Rees1984,NiemJed1999,CamposLima2012}, and so on. The leading roto-translational effects in deflection of photon and particle presented above may provide some possibilities to study them the other way around, although the measurement of kinematic effects may be difficult~\cite{Sereno2007}.

Since the existence of gravimagnetism (extrinsic gravimagnetic field) has been verified preliminarily by Jovian deflection experiment~\cite{FomKopeikin2003}, perhaps the coupling effects of translational motion and rotation can be also used to investigate it. As demonstrated in previous works (see, e.g., Refs. \cite{KopeiFoma2007,Kopeikin2007}), the translational motion of Kerr black hole can bring about extrinsic gravimagnetic field while the rotational-type mass currents can induce intrinsic gravito-magnetism. Note that intrinsic gravimagnitic field cannot be removed using coordinate transformation~\cite{Sereno2003,Sereno2005b} and that two types of gravimagnetic field can be probed via time delay and/or relativistic deflection of light, respectively~\cite{Sereno2005a,Kopeikin2006,KopeiFoma2007}. Therefore, the persistence of the coupling of the motions reflects the coupling between the intrinsic and extrinsic gravito-magnetic field always persists. This coupled gravimagnetic field is also showed by the explicit metric of moving Kerr black hole. And the analyses of the influence of roto-translational effects on the contribution of SOKD in Section~\ref{rotational effect} and the explicit formula such as Eqs.~\eqref{angle2} and \eqref{angle4} might be helpful to study them.

\section{discussion and summary} \label{discussion}
Kinematic effects in the first-order Schwarzschild deflection of the test particles were studied systematically~\cite{Olaf2004}. In this paper, we investigate the roto-translational effects on the SOKD in the weak gravitational field of one uniformly moving Kerr black hole. We analytically calculate the contribution of the kinematically corrected SOKD to the total deflection angle, and obtain its upper bound. The dependence of this ratio on roto-translational effects has been dealt with from three sides.

For the critical velocity $w=1/\sqrt{3}$ of test particle in the limit of FOV approximation, the translational effects always persist for corrected SOKD even though they vanish in the first-order deflection. In the limit of $v\rightarrow 1$, if we take into account of the kinematically corrected second-order Schwarzschild deflection which reduces to $\frac{15\pi m^2}{4b^2}$ for light deflection by nonmoving Kerr black hole (see, e.g., Ref. \cite{EpsShap1980}), the total second-order bending angle may not disappear.

In comparison with the work of Sereno~\cite{Sereno2005b}, we have employed Lagrange equations to study the roto-translational effects on the deflection of both photon and massive particles. At the same time, our results are not limited to the low velocity of deflector.

Based on the strict harmonic Kerr metric and Lorentz transformation, one can extend this method to calculate the velocity or roto-translational effects on the deflection to arbitrary order, in principle. The effects of transversal velocity can also be investigated. With the metric in the weak-field limit, i.e., Eqs. \eqref{g00m}\,-\,\eqref{gijm}, we can also study other observable effects such as the Shapiro time delay and redshift of light in the time-dependent gravitational field of the moving Kerr black hole, and this work will be presented in next paper.

\section*{Acknowledgments}
This work was supported in part by the Program for New Century Excellent Talents in University (No. NCET-10-0702), the National Basic Research Program of China (973 Program) Grant No. 2013CB328904, and the Ph.D. Programs Foundation of Ministry of Education of China (No. 20110184110016).

\end{document}